*Original Article*

# Study on Standardizing Working Time: A Case of XYZ Retail Store in Bandung, Indonesia

[1]Aprodhita Anindya Putri, [2]Akhmad Yunani
[1,2] School of Communication and Business, Telkom University, Jl. Telekomunikasi No. 1 Ters. Buah Batu, Bandung, Indonesia

***Abstract:*** *Work time standardization helps to find and reduce wasteful movements and time in the workplace, such as chatting, mobile phone use, insufficient rest, or unproductive tasks. This study aims to map the process of displaying products from the warehouse to the shelves and calculate and determine the standard working time of employees of the Operations Division of PT XYZ Branch who oversee displaying X Milk and Y Bread. The data was collected six times in three weeks including interviews and observations and took a sample of 20 pieces on each product to carry out data analysis such as data sufficiency tests and control charts. Several time deviations were found in the display process of X Milk products on all observation days in different activities. Whereas in the process of displaying Y Bread, only the deviation of working time was found on the 4th observation day, which proves that the process needs to have a standard working time so that the activity work time is more controlled. Therefore, the analysis is carried out with the calculation of performance rating, time allowance, normal time, and standard time. The result of the standard time calculation for the display process of X Milk products is 15.83 minutes and Y Bread is 9.18 minutes for each product of 20 units.*

## I. INTRODUCTION

The intense competition in the retail industry makes companies strive to meet customer needs in terms of specifications, quantities, and time requested and at the lowest cost. In the retail industry, in addition to inventory turnover and layout optimization, employee standard working time is an important issue that impacts the performance of retail companies. Research on the standardization of supermarket employees' work time is a relevant and useful topic to address issues facing the industry. Standardizing performance time helps to find and reduce wasteful movements and time in the workplace, such as cell phone use, insufficient breaks, or unproductive tasks. According to Gustafsson (2019), waste of motion is the movement of employees or machines that do not provide added value to customers, but only adds cost and time to the production process.

A precise measurement technique is essential to give precise information on the amount of time needed and the effectiveness of the movement of each activity, given the significance of monitoring every task completed by personnel. Labor or employees are the most important of all these elements because the workforce will run the wheels of the business. Labor has different abilities, ranging from the very proficient to the mediocre, and there are even those who are very slow or below the average of other employees. The tight competition between companies makes companies must strive to meet customer needs in terms of specifications, quantities, and time requested and at the lowest cost, where ultimately, employee performance capability is a very important component for operational efficiency. Research on the standardized working time of supermarket employees is a relevant and useful topic to address the issues facing the industry.

Work time standardization helps to find and reduce wasteful movements and time in the workplace, such as chatting, mobile phone use, inadequate breaks, or unproductive tasks. Recognizing the value of measuring each task the business performs, a precise measurement technique is necessary to give exact data on the amount of time needed and the effectiveness of each task's flow to generate goods. Standardized work can help retail businesses improve customer service in several ways. It can ensure consistency in product and service quality, which is a prerequisite for customer satisfaction. Standardization of working time is an effective way to improve a process. Standardized work provides optimal consistency and prevents quality degradation in stores. Standardized is most often the most well-known way to complete a particular task, making it the safest and most effective way to ensure that deliveries are made on time, neatly, and to a high standard.

One of the methods used in measuring the time of the activities used is by using the Time and Motion Study (TMS) method. In this case, the TMS method has become an important tool for analysis, understanding, and improving employee performance efficiency. Frederick Winslow Taylor first developed this method in the early 20th century as part of the scientific management movement.

According to Khandve (2017), TMS or time and motion study is a brand-new integrated methodology that uses a variety of work-study techniques to increase efficiency in the construction industry. The research on productivity enhancement employed a work-study methodology in conjunction with contemporary soft skills. By applying temporal motion studies or TMS to improve employee planning, human performance was evaluated. It was shown that the time taken to perform work effectively could be reduced by 30 percent, thereby reducing project cost overruns. Modern industry 4.0 technology could revolutionize time studies soon. XYZ Store is an Indonesian original modern retail company with Supermarket and Department Store formats that has large operational costs, especially labor costs requiring standardization of performance time to help control costs by ensuring employees work efficiently according to their duties and responsibilities. In this case, employees in the Operations Division of XYZ Store who have the task of displaying products or displaying products from the warehouse to the shelves certainly perform many effective and ineffective movement activities. Movement activities that are considered less or even ineffective often result in less-than-optimal work results for the employees themselves.

In the preliminary research that has been carried out temporarily, there are several core or main activities in the process flow of displaying products to the shelves, such as carrying products from the warehouse to the shelves →, tidying up the shelves → displaying products to the main shelves and additional shelves (additional shelves rented by product suppliers). There are several other activities where employees perform activities that are not the main or not the core of the job, such as re-lowering items that are already on the shelf as a whole and wiping products extra. In this case, the decision to eliminate the activity is clearly not possible because the activity aims to satisfy consumers, who are the target of this business. So, the right solution to this phenomenon is a good standard working time for maximum work output and time. The goal of standard time is to identify and subsequently remove motion and time inefficiencies in order to create standardized and efficient processes for carrying out activities and to gauge the performance of operators. Standard time can eliminate waste and inconsistency of work. No major changes have occurred; standardizing work performed in any field in the most effective manner remains a crucial component of increasing efficiency. TMS is the study of the motions employees use to finish their tasks. With this study, it is intended to obtain standard movements for the completion of a job, namely a series of effective and efficient movements. The application of TMS to XYZ Store employees is a method for process improvement and development of labor standards. Although Standard Operating Procedures (SOP) have been established, currently, XYZ Store does not have a standard working time for the Branch Operations Division. It is feared that the absence of standard working time will result in a decrease in employee productivity, which in turn will reduce the company's business performance. This research also involves the observation and analysis of time and motion in various tasks and processes in a retail environment. This resulted from initial observations for approximately two-hour periods in one week, which showed that employees who performed the process of displaying products to shelves in each activity had inconsistent times.

Research on standardization of working time for employees of the Operational Division of this branch of XYZ Store is expected to improve supermarket profitability and find practical recommendations that can help supermarkets improve operational efficiency, optimize the use of human resources, and achieve long-term success in a competitive environment.

## II. LITERATURE REVIEW

*A) Standard Working Time*

Standard working time, according to Freivalds and Niebel (2009), is the time required to complete a work process. The results of determining standard time and standard output become the benchmark used to assess the performance of employees and become an assessment of the understanding of work procedures and the ability to complete their work. Standard time can be used to indicate how long activities should take, how much output will be produced, and how much labor is needed to complete the work (Zadry et al., 2015). According to Muhacır et al. (2022), there has been no significant change from the past until now; the standardization of work done in any field done in the best way is a very important element in improving efficiency. According to Lukodono and Ulfa, (2017), standardized time can eliminate waste and inconsistency of work.

*B) Waste of Motion*

Beginning with research conducted by Frank Gilbreth and his wife, Lillian Gilbreth, on the basic movements made by humans while working, which later became known as motion studies, the results showed that the movements or activities found were unnecessary or called waste (Zadry et al., 2015). According to Gustafsson (2019), waste of motion is the movement of employees or machines that do not provide added value to customers. But only adds cost and time to the production process.

*C) Time Study*

According to Ralph M. Barnes (1980), a time study calculates how long it takes a qualified person to work at a normal pace to complete a particular task. Time study is an activity to determine the time required by employees to do their work under normal conditions. The purpose of a work measurement system is to determine the average time required to do a job by a trained employee doing a job. This time is called standard time. According to Permata & Hartanti (2016), a job is considered to be completed efficiently if the completion time lasts the shortest. The method is used because it can determine the standard time by considering fatigue and delays. Time study is also an aspect that consists of a variety of procedures to determine the length of time required with a standard time measurement set for each activity involving humans, machines, or a combination of activities.

*D) Motion Study*

According to Mundel & Danner (1994), the term of motion study is an aspect consisting of description, systematic analysis, and development of work methods in determining raw materials, output design, processes, work tools, workplaces, and equipment for each step in a process or human activity that does each activity itself. The purpose of the motion study method is to determine or design a suitable work method to complete an activity.

*E) Time and Motion Study (TMS)*

According to Bon & Daim (2010), Time and Motion Study (TMS) has the aim of eliminating unnecessary work. Motion studies mostly focus on analysis while time studies involve more measurement. The result of the TMS method, then, is the number of minutes it would take for a person who is suited to the job and trained in the prescribed ways to complete it if they were working at normal times. This time is known as the standard operating time. According to Nurdiansyah and Satoto (2023) TMS is a study of the movements performed by workers to complete their work. With this study, it is intended to obtain standard movements for the completion of a job, namely a series of effective and efficient movements. Meanwhile, according to Barnes (1980), TMS is a systematic study of work systems with the aim of creating better systems and methods, setting standards for systems and methods, and helping train employees to apply better methods.

*F) Data Sufficiency Test*

The data sufficiency test is used to see whether the data taken is statistically sufficient or not. In this test, a confidence level of 95% and a degree of precision of 5% were used. The calculation of data sufficiency uses the following equation:

$$N' = \left[\frac{k/s\sqrt{N(\Sigma xi^2) - (\Sigma xi)^2}}{(\Sigma xi)}\right]$$

N'= Number of samples required
N = Number of samples taken
xi= Observation data
s = Degree of precision. If the confidence level is 95%, then s = 5% k = 95% confidence level with k = 2 which shows the amount of confidence of the measurer in the accuracy of Anthropometric data

If N' ≤ N, then the amount of data is sufficient. If N' > N, then the amount of data is not enough.

*G) Control Diagram*

Control charts are used to test the consistency of existing time study data. A control diagram is a tool to analyze is data controlled or not controlled from observations. Data that is between the two control limits means that it is controlled, otherwise if the data is outside the control limits, it means that there is a deviation in the data.

*H) Performance Rating*

Performance rating or level of work on employees to see the time equation of the results of observations of employees in completing a job using the time required by normal employees in completing the job (Freivalds and Niebel, 2009). One way to determine the performance rating is using the Westing House method which consists of four factors that affect the rating value such as skill, effort, working conditions, and consistency (Permata and Hartanti, 2016). The performance rating according to the Westing House Method can be depicted in Table 1 below.

Table 1: Performance Rating According to the Westing House Method

| Skill | Effort | Conditions | Consistency |
|---|---|---|---|
| Super<br>A1= + 0,15<br>A2= +0,13 | Excessive<br>A1= +0,13<br>A2= +0,12 | Ideal<br>A= +0,06 | Perfect<br>A= +0,04 |
| Excellent<br>B1= +0,11<br>B2+ +0,08 | Excellent<br>B1= +0,10<br>B2= +0,08 | Excellent<br>B+ +0,04 | Excellent<br>B= +0,04 |
| Good<br>C1= +0,06<br>C2= +0,03 | Good<br>C1= +0,05<br>C2= +0,02 | Good<br>C= +0,02 | Good<br>C= +0,01 |

| **Skill** | **Effort** | **Conditions** | **Consistency** |
|---|---|---|---|
| Average<br>D= 0,00 | Average<br>D= 0,00 | Average<br>D= 0,00 | Average<br>D= 0,00 |
| Fair<br>E1= -0,05<br>E2= -0,10 | Fair<br>E1= -0,04<br>E2= -0,08 | Fair<br>E= -0,03 | Fair<br>E= -0,02 |
| Poor<br>F1= -0,16<br>F2= -0,22 | Poor<br>F1= -0,12<br>F2= -0,17 | Poor<br>F= -0,07 | Poor<br>F= -0,04 |

*I) Normal Time*

Calculation Normal time is defined as the amount of time required by an operator who has average skills to perform tasks in normal time and conditions. Normal time can be obtained using the following formula: Cycle time is obtained from the average of five observation time data for each process on each product. The calculation of normal time uses the following formula:

$$Nt = \text{Cycle Time} \times \text{Performance Rating (Rating Factor)}$$

Where:

Cycle Time= average obtained from all studies

Performance rating= the sum of four factors that affect the rating value such as skills, effort, working conditions, and consistency according to the Westing House Table.

*J) Standard Time*

Calculation Standard time is calculated using normal time that considers the condition of workers. The method for calculating standard time is as follows:

$$St = Nt \times \frac{100\%}{100\% - \text{allowance}\%}$$

Allowance= Time allowance resulting from calculations of:

$$\frac{\text{Total minutes of break time}}{\text{working time}} \times 100\%$$

## III. RESULTS AND DISCUSSION

*A) Movement Analysis through Product Display Process Map*

Movement analysis is an activity of observing what movements employees do in working on a process that will be further researched with the TMS method. Movement analysis is carried out using a process map creation. The following is an image of the flow map of the X Milk and Y Bread product display process, which consists of several activities or sub-processes:

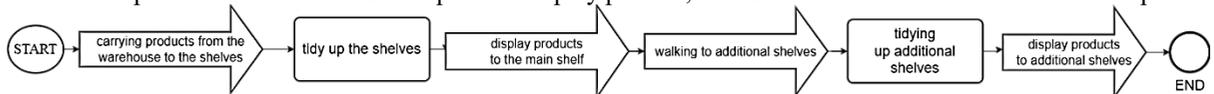

**Figure 1:** X Milk Display Process Map

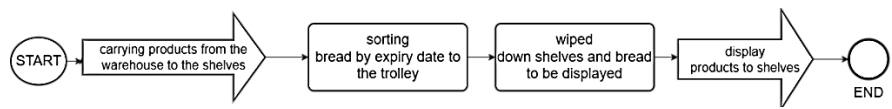

**Figure 2:** Y Bread Display Process Map

*B) Data Collection Time*

Calculation Based on the results of five observations or observations of the calculation of the process of displaying or displaying products to shelves which have several activities on X Milk and Y Bread products, the results of time calculations with a stopwatch time measuring device are obtained as follows:

Table 2: Data Collection Results Calculation of X Milk Display Process Time

| Activities | Observed time | | | | | Total |
|---|---|---|---|---|---|---|
| | 1 | 2 | 3 | 4 | 5 | |
| carrying products from the warehouse to the shelves | 1,58 | 1,45 | 1,52 | 1,7 | 1,58 | 7,83 |
| tidy up the shelves | 1,8 | 1,95 | 2,1 | 1,85 | 1,98 | 9,68 |
| display products to the main shelf | 2,67 | 2,88 | 2,87 | 2,75 | 3,05 | 14,22 |
| walking to additional shelves | 1,49 | 1,37 | 1,3 | 1,32 | 1,29 | 6,77 |
| tidying up additional shelves | 1,58 | 1,73 | 1,57 | 1,48 | 1,55 | 7,91 |
| display products to additional shelves | 2,83 | 3,03 | 2,72 | 2,7 | 3,07 | 14,35 |

Table 3: Data Collection Results Calculation of Y Bread Display Process Time

| Activities | Observed time | | | | | Total |
|---|---|---|---|---|---|---|
| | 1 | 2 | 3 | 4 | 5 | |
| carrying products from the warehouse to the shelves | 0,84 | 0,83 | 0,83 | 0,95 | 0,87 | 4,32 |
| sorting bread by expiry date to the trolley | 2,49 | 2,44 | 2,43 | 2,80 | 2,61 | 12,77 |
| wiped down shelves and bread to be displayed | 1,15 | 1,17 | 1,17 | 1,33 | 1,23 | 6,05 |
| display products to shelves | 2,32 | 2,28 | 2,33 | 2,64 | 2,47 | 12,04 |

*C) Data Sufficiency Test*

The X Milk product display process is one of the processes studied and the data sufficiency test is calculated. Where in the process there are six activities. The first activity that was tested for data adequacy was the activity of Carrying Products from the Warehouse to the shelf as below:

Table 4: Sufficiency Test of Road Activities Carrying X Milk from the Warehouse to the Shelf

| Acivity | Observed time | | | | | Total |
|---|---|---|---|---|---|---|
| carrying products from the warehouse to the shelves | 1 | 2 | 3 | 4 | 5 | |
| X | 1,58 | 1,45 | 1,52 | 1,70 | 1,58 | 7,83 |
| X² | 2,51 | 2,10 | 2,30 | 2,89 | 2,51 | 12,31 |

The data sufficiency test for the activity of carrying products from the warehouse to the shelves is calculated as follows:

$$N' = \left[\frac{k/s\sqrt{N(\Sigma xi^2) - (\Sigma xi)^2}}{(\Sigma xi)}\right]$$

The initial stage that needs to be done is to find the value of $k/s$ first where $k = 2$, and $s = 0.05$. Then proceed to calculate the value of $(\Sigma xi)^2$ and $(\Sigma xi^2)$ resulting from the calculations below:

$$(\Sigma xi)^2 = (1,58+1,45+1,52+1,70+1,58)^2$$
$$= 7,83^2$$
$$(\Sigma xi^2) = (1,58^2+1,45^2+1,52^2+1,70^2+1,58^2)$$
$$= 12,31$$

Then the value of N' can be known as follows:

$$N' = \left[\frac{40\sqrt{5(12,31) - (7,83)^2}}{(7,83)}\right]$$
$$N' = 4,49$$

The result of calculating the value of N' = 4.49 is smaller than the number of observations that have fulfilled the rule of N' < N, namely 4.49 < 5. Thus, the data for the process of displaying X Milk products in the activity of Carrying Products from the Warehouse to the Shelf is sufficient.

So, based on calculations that have been carried out carefully, a recapitulation of the results of the data adequacy test of the process of displaying X Milk and Y Bread are depicted in the following tables:

**Table 5: Recapitulation of Data Sufficiency Test Results for the X Milk Display Process**

| Activities | N' Value | Description |
|---|---|---|
| carrying products from the warehouse to the shelves | 4,49 | Sufficient |
| tidy up the shelves | 4,62 | Sufficient |
| display products to the main shelf | 3,36 | Sufficient |
| walking to additional shelves | 4,74 | Sufficient |
| tidying up additional shelves | 4,33 | Sufficient |
| display products to additional shelves | 4,63 | Sufficient |

**Table 6: Recapitulation of Data Sufficiency Test Results for the Y Bread Display Process**

| Activities | N' Value | Description |
|---|---|---|
| carrying products from the warehouse to the shelves | 4,42 | sufficient |
| sorting bread by expiry date to the trolley | 4,57 | sufficient |
| wiped down shelves and bread to be displayed | 4,98 | sufficient |
| display products to shelves | 4,84 | sufficient |

Based on Tables 5 and Table 6 above, it is concluded that conducting research or observation five times in the process of displaying X Milk and Y Bread products produces sufficient data and can be processed to the next stage of calculation.

### D) *Control Diagram*

The time results for each activity in the flow were classified into subgroups to obtain the meantime, standard deviation, UPL, and LCL. Control charts were used to test the consistency of the time study data regarding the presence or absence of time deviations in performing these activities, as depicted in Table 7 and Figure 3 below.

**Table 7: Data Result of Average, Standard Deviation, UPL, and LCL Calculation of Walk Activity from Warehouse to the X Milk's Shelf**

|  | Observed Time | Average | STD | UPL | LCL |
|---|---|---|---|---|---|
| 1st Observation | 1,58 | 1,57 | 0,09 | 1,66 | 1,47 |
| 2nd Observation | 1,45 | 1,57 | 0,09 | 1,66 | 1,47 |
| 3rd Observation | 1,52 | 1,57 | 0,09 | 1,66 | 1,47 |
| 4th Observation | 1,7 | 1,57 | 0,09 | 1,66 | 1,47 |
| 5th Observation | 1,58 | 1,57 | 0,09 | 1,66 | 1,47 |

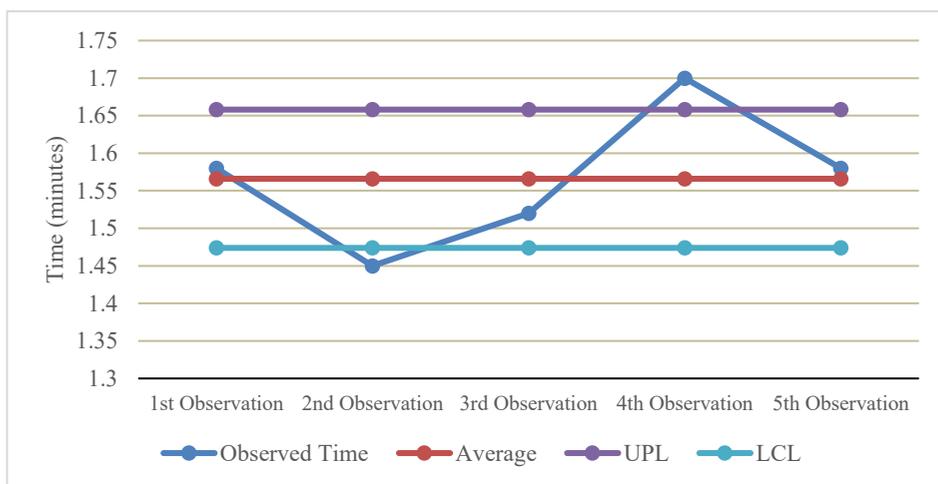

**Figure 3: Activity Control Diagram for Carrying X Milk from the Warehouse to the Shelf**

The existence of one point outside the upper limit of the yellow line, which is the UPL line, and one point outside the lower limit or blue line, which is the LCL line, shows that there are time deviations encountered from five studies in the same activity, namely on the 2nd and 4th research days. This is due to the absence of standard time in carrying out the product display process at the Branch Operations Division at PT.XYZ.

The discovery of deviations that occur in each activity with the research day on various days is thought to be related to the lack of clarity of rules that make employee performance inconsistent so that this becomes a new finding that time standards are needed by employees in the Branch Operations Division so that on any day and in any situation, employees will remain consistent with regular working time.

*E) Rating Factor*

The calculation of worker performance or rating factor is using the Westinghouse Table which consists of skill, effort, condition, and consistency factors. The rating factor under normal circumstances is +1. The rating factor value used is assumed to be the same for all processes and all products. Based on the four factors from Table 1, the skill and effort factors are in the Good (C1) category and the condition and consistency factors are in the Good (C1) category. So, the rating factor value is obtained, namely:

$$1 + (0,06+0,05+0,02+0,01) = 1.14.$$

*F) Normal Time Calculation*

Normal time is the time with conditions or conditions and normal work speed required by average skilled employees in doing work. The search for normal time cannot be separated from cycle time. Cycle time is obtained from an average of five observation time data for each process on each product. The calculation of normal time uses the following formula:

$$Nt = Cycle\ Time \times Performance\ Rating\ (Rating\ Factor)$$

The calculation below is the normal time of the activity of Carrying Products from the Warehouse to the Shelf in the X Milk product display process:

$$Nt = 1.56 \times 1.14 = 1.79\ minutes$$

The results of the above calculations show that employee activities in the form of Walking from the Warehouse to the Shelf take 1,785 minutes. Then, Table 8 and Table 9 below are the recapitulations of the results of the calculation of normal time for each activity in the product display process of X Milk and Y Bread:

**Table 8: Recapitulation of the Normal Time Calculation Results of the X Milk Display Process**

| | Activities | Cycle Time (minutes) | *Rating Factor* | Normal Time (minutes) |
|---|---|---|---|---|
| X Milk | carrying products from the warehouse to the shelves | 1,57 | 1,14 | 1,79 |
| | tidy up the shelves | 1,94 | 1,14 | 2,21 |
| | display products to the main shelf | 2,84 | 1,14 | 3,24 |
| | walking to additional shelves | 1,35 | 1,14 | 1,54 |
| | tidying up additional shelves | 1,58 | 1,14 | 1,80 |
| | display products to additional shelves | 2,87 | 1,14 | 3,2718 |

**Table 9: Recapitulation of the Normal Time Calculation Results of the Y Bread Display Process**

| | Activities | Cycle Time (minutes) | *Rating Factor* | Normal Time (minutes) |
|---|---|---|---|---|
| Y Bread | carrying products from the warehouse to the shelves | 0,86 | 1,14 | 0,98 |
| | sorting bread by expiry date to the trolley | 2,55 | 1,14 | 2,91 |
| | wiped down shelves and bread to be displayed | 1,21 | 1,14 | 1,38 |
| | display products to shelves | 2,40 | 1,14 | 2,75 |

G) *Allowance Calculation*

In calculating standard time, allowance is needed. Allowance is the total free time set such as time for personal needs, relieving fatigue (fatigue), and unavoidable things. Merrick suggested that fatigue and other allowances should be added to the measured times of job. Personal needs include drinking when thirsty, going to the restroom, and chatting with colleagues. Reported by an interview with the Supervisor of the Branch Operations Division of PT XYZ, the working hours of employees in the division are 8 hours on each shift. While the rest time given by the company is for one hour. The following is a breakdown of the percentage value of leeway in the production process at PT. XYZ:

Working time = 8 hours x 60 minutes = 480 minutes
Break time= 60 minutes
Then, the following leeway time is generated:

$$\frac{60}{480} \times 100\% = 12{,}50\%$$

Based on the calculation of the leeway time that has been carried out, it is known that the leeway time required by employees in the process of displaying X Milk and Y Bread products at PT XYZ is 12.50%. The leeway time that has been obtained will later be used to calculate the standard time.

H) *Standard Time Calculation*

The results of normal time and allowance time that have been found previously can be directly used to calculate the standard time of the first activity, namely the activity of carrying products from the warehouse to the Shelf on X Milk products as follows:

$$St = 1{,}79 \times \frac{100\%}{100\% - 12{,}50\%}$$

$Ws$ = 2,04 minutes for 20 units

The result of the calculation of standard time on the activity of carrying products from the warehouse to the shelf on X Milk products is 2.04 minutes for 20 units. This means that employees have an average ability to complete an activity in 2.04 minutes.

Furthermore, Table 10 and Table 11 below are the recapitulations of the results of the calculation of the standard time of each activity in the X Milk and Y Bread product display process:

Table 10: Recapitulation of Standard Time Calculation Results of X Milk Display Process

| | Activities | Normal Time (minutes) | Allowance | Standard Time (minutes) |
|---|---|---|---|---|
| X Milk | carrying products from the warehouse to the shelves | 1,79 | 12,50% | 2,04 |
| | tidy up the shelves | 2,21 | 12,50% | 2,52 |
| | display products to the main shelf | 3,24 | 12,50% | 3,71 |
| | walking to additional shelves | 1,54 | 12,50% | 1,76 |
| | tidying up additional shelves | 1,80 | 12,50% | 2,06 |
| | display products to additional shelves | 3,27 | 12,50% | 3,74 |
| | | | Total | 15,83 |

Table 11: Recapitulation of Standard Time Calculation Results of Y Bread Display Process

| | Activities | Normal Time (minutes) | Allowance | Standard Time (minutes) |
|---|---|---|---|---|
| Y Bread | carrying products from the warehouse to the shelves | 0,98 | 12,50% | 1,13 |
| | sorting bread by expiry date to the trolley | 2,91 | 12,50% | 3,33 |
| | wiped down shelves and bread to be displayed | 1,38 | 12,50% | 1,58 |
| | display products to shelves | 2,75 | 12,50% | 3,14 |
| | | | Total | 9,18 |

## IV. CONCLUSION

Based on research on working time at PT XYZ approximately six times in three weeks as well as analysis and processing of existing data, it can be concluded as follows.

1. The process of displaying X Milk products begins with walking from the warehouse to the shelves, tidying the shelves, and displaying the products. Because the supplier rents another shelf as an additional shelf that is placed in a different aisle from the main shelf, employees must walk towards the shelf to tidy up and display the products on the shelf.
2. Bread product display process also begins with walking the products from the warehouse to the shelves followed by sorting the bread based on the expiry date to the trolley, wiping the shelves and bread to be displayed, and displaying the products to the shelves.
3. The standard time for the display process of X Milk products is 15.83 minutes for 20 products, including the activities of walking carrying the product to the shelf for 2.04 minutes, tidying the shelf for 2.52 minutes, displaying the product to the shelf for 3.71 minutes, walking to the additional shelf for 1.76 minutes, tidying the additional shelf for 2.06 minutes, and displaying the product to the additional shelf for 3.74 minutes.
4. The standard time for the display process of Y Bread product display process is 9.18 minutes for 20 units, including the activities of walking carrying products from the warehouse to the shelves with a standard time of 1.13 minutes, sorting bread based on its expiry date to the trolley for 3.33 minutes, wiping the shelves and bread to be displayed for 1.58 minutes, and displaying products to the shelves for 3.14 minutes.